# DIALOG: A framework for modeling, analysis and reuse of digital forensic knowledge

Damir Kahvedzic´*, Tahar Kechadi1

Center for Cybercrime Investigation, University College Dublin, Ireland

**Abstract**

This paper presents DIALOG (Digital Investigation Ontology); a framework for the management, reuse, and analysis of Digital Investigation knowledge. DIALOG provides a general, application independent vocabulary that can be used to describe an investigation at different levels of detail. DIALOG is defined to encapsulate all concepts of the digital forensics field and the relationships between them. In particular, we concentrate on the Windows Registry, where registry keys are modeled in terms of both their structure and function. Registry analysis software tools are modeled in a similar manner and we illustrate how the interpretation of their results can be done using the reasoning capabilities of ontology.

**Keywords:**

Windows, Registry, Digital, Investigation, Ontology

## 1. Introduction

The rate of computer crime continues to increase year to year. The sophistication of the crimes and the variety of technological devices employed in these offenses are becoming critical challenges to the investigators (Sophos, 2009; U.S. Department of Justice, 2007). As well as the inherently distributed cyber crimes, such as DOS attacks, low-level cyber crimes involving only a few individuals now typically involve the investigation of multiple devices. Consequently digital investigations are more prolonged, complicated and require the integration of many disparate sources of data.

As a result investigators require extensive training in a wide range of software tools, techniques, hardware equipment and digital devices. In addition to being aware of emerging technologies and possible sources of evidence, investigators need also be aware of inaccuracies and applicability of using a particular technique or tool on a particular device for a particular case.

Guides are continually being published to advise the investigators on how to investigate a particular device and carry out the investigation effectively (Carvey, 2005; Farmer and Burlington, 2007; Sophos, 2009; U.S. Department of Justice, 2007; Wong, 2009). The transfer of knowledge to relevant parties is informal and periodic. A central, application and case independent knowledge base that can be continually supplemented with new knowledge can be an invaluable resource of reference to an investigative team. The knowledge base, designed in a logic manner, would reflect the digital forensic field and give structure to an investigation by defining each of the main major concepts and their attributes.

Ontologies have been developed for the Semantic Web to give a structure to the seemingly unstructured world of the Internet. They are a ''formal, explicit specification of shared conceptualisation'' (Gruber, 1995) providing a vocabulary to model various domains. They have diversified to model such domains as biomedicine (The Open Biomedical Ontologies, 2009) and everyday common sense knowledge (Cycorp Inc., 2009). Full ontology languages, restrictions and rules have been developed to work only on this meta information and allow models to infer new knowledge.

In this paper we present the *Digital Investigation Ontology*, DIALOG, an ontology for the representation, reuse and analysis of Digital Investigation knowledge. DIALOG contains the main concepts of digital forensics and their relationships and captures the universe of discourse of the Digital Investigation domain. It is designed to be independent of any specific investigation and can grow by progressively expanding its domain knowledge with definitions of new entities in a similar way to other ontologies.

DIALOG is envisioned to play a number of roles in the Digital Investigation field:

1) *As a knowledge repository:* DIALOG can be instantiated with specific pieces of information that can be searched for by investigators if they encounter it in a case and do not know what it is.
2) *As a case manager:* Evidence relating to a specific case can be annotated in DIALOG and provide a central place where information can be shared between relevant parties, therefore facilitating collaboration.
3) *As an evidence unification mechanism:* Similar to the above, evidence from different devices can be annotated and rules can be employed to resolve logic inconsistencies that may arise.
4) *As an investigation guide:* As well as definitions and conceptualisation, DIALOG can include warnings, metrics and other abstract concepts to guide the investigator away from making mistakes.

We will limit the scope of this paper to the encoding of forensics knowledge associated with the Windows Registry. The Registry is a central database storing a vast amount of information about the system resources (software and hardware), its users and their preferences. Guides, similar to those of the file system, have been published to analyse the registry (Registry Hives, 2008). The scope of the evidence held within it, its importance in the investigation and the wide variety of tools available for its analysis make it analogous to the file system. Expansion of DIALOG to represent information with respect to the file system can therefore be achieved in the same way as for the registry. Modeling other specific areas of interest can also be incrementally added to the ontology.

In particular, we will use the registry to illustrate DIALOG's role as a knowledge repository and case manager (points 1 and 2 above). It is unlikely that investigators would be familiar with all registry keys and the purposes. We will illustrate how DIALOG can model the registry key and serve as a reference for unfamiliar keys. As a case manager, DIALOG can annotate evidence from existing investigation tools to add meaning to the results. We enhance the registry analysis software, RPCompare (Kahvedžić and Kechadi, 2008). Using DIALOG, RPCompare can annotate its results and use formal rules to interpret them and automatically classify the evidence into categories. The rules can be checked and verified for consistency using existing ontology logic reasoning.

Section 2 describes DIALOG. The ontology is an expressive entity and can elaborate and refine the definitions of the digital investigation concepts by relating them to each other across branches of a taxonomic tree. DIALOG consists of four main sub-ontologies discussed in detail in Section 2.

In Section 3, we discuss the use of various sub-ontologies and concepts to model the knowledge associated with the Windows Registry. RPCompare is modeled with respect to both its structure and its operations in Section 4. The advantage of using an application independent model to manage the results of RPCompare is discussed. In particular, we illustrate the ability of DIALOG to annotate the results with evidence concepts and infer new knowledge. Section 5 concludes the paper.

## 2. DIALOG framework

An ontology is an abstract description of concepts and their relationships in a given universe of discourse. It creates a formal, application independent vocabulary that can be reused across different fields. Currently, the ontology models the digital forensics field through four main dimensions.

- *Crime Case*: Types of investigations based on the crime suspected to have been committed.
- *Evidence Location*: Types of locations or sources of evidence that can be searched to find evidence.
- *Information*: Types of information (files, software) that can be found in the system.
- *Forensic Resource*: Types of resources (tools, software) that can be employed to carry out an investigation.

The *Crime Case*, *Evidence Location* and *Information* ontologies are orthogonal to each other and define distinct concepts and entities of the domain. The *Forensic Resource* ontology, on the other hand, can be viewed as a *specialisation* of the *Information* ontology. It defines tools and other concepts used specifically in the forensic field and is in fact mirrored in the relevant place in the *Information* ontology.

Fig. 1 shows the hierarchy of the top level concepts of the sub-ontologies. The knowledge base is constructed by creating instances of the concepts with their relevant relations and restrictions. The figure omits more specific concepts and only shows the *is_a* relations between them. More detailed description of the sub-ontologies including their relations are shown in subsequent sections.

### 2.1. Crime case ontology

Every case starts by setting an aim, namely to prove or disprove if one or more crimes have occurred, and no investigation can be carried out if no crime is suspected to have happened. The *Crime Case* ontology is the main ontology for description of cases and catagorises different investigation types in terms of the suspected crime. Since the *Crime Case* and *Crime* concepts are analogous, *Crime* taxonomies (JISC Legal, 2007; Shinder, 2002; U.S. Department of Justice, 2008) are used as starting points in developing the *Crime Case* ontology. An investigation may fall into one or more *Crime Case* category if one or more suspected crimes are present.

There are a variety of ways that crimes and investigations can be organised by an ontology. Computers can be used as a target or tool to commit high tech versions of crimes that have evolved out of the traditional non digital realm, such as the *Fraud* and *Extortion*, or simply contain supporting evidence of inherently non digital crimes such as *Murder*. The ontology defines the *NonCyberCrimeCase* and the *CyberCrimeCase* as the two most general and disparate concepts to differentiate investigations between these two types of crimes.

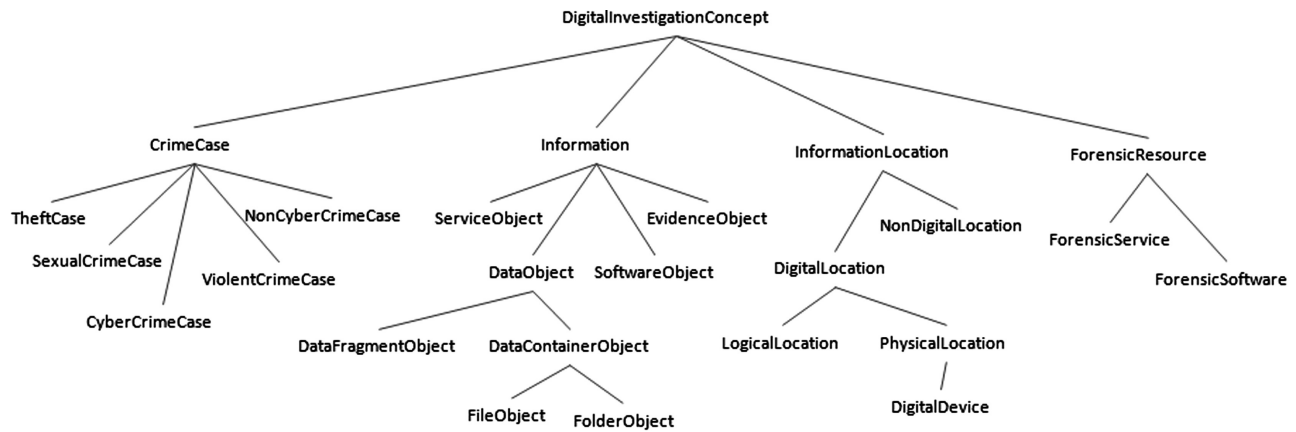

Fig. 1 – Top Level of the Ontology.

The *NonCyberCrimeCase* concept conceptualises those investigations of crimes that can *never* be conducted in the digital world, such as *Murder*, and which happen to have evidence in a digital form. The ontology does not provide an exhaustive characterisation of them. A small number of concepts, such as the *HomicideCase*, *DomesticViolenceCase*, and *KidnapCase* concepts are defined since they have been discussed in other digital investigation guides (U.S. Department of Justice, 2008). The *CyberCrimeCase*, on the other hand, defines those investigations of crimes that have a definitive digital component. In *CyberCrimesCase*, the evidence found on the computer, either the data stored in it or the user actions carried out with it constitute a crime.

Three top level concepts are defined in the ontology to differentiate between *all* types of investigations. The *TheftCase*, *ViolentCrimeCase* and *SexualCrimeCase* concepts are used as container classes to generalise the wide variety of cases. The number of these has been kept to a minimum and expresses the domain more accurately. The *CyberTheftCase* (theft in a digital environment) for example, is defined as both *TheftCase* (crimes that involve unlawful appropriation) and *CyberCrimeCase*. Similarly, *CyberFraudCase* is both *FraudCase* (theft by deception) and *CyberCrimeCase*.

*DigitalMaterialCrimeCase* conceptualises all investigations of crimes that are perpetrated if a person possesses or propagates content that has been deemed illegal. These *PropagationOfUnlawfulMaterialCase* and *PossessionOfUnlawfulMaterialCase* concepts differ from *TheftCase* in that these materials are not assumed to be stolen. The concepts are disjoint since it may be lawful for a person to possess something but unlawful for them to distribute it, such as copyrighted material. On the other hand, it is illegal to posses child pornography even if the suspect does not distribute it.

All other crimes fall into one or more of the following categories: *CyberTheftCase*, *CyberFraudCase*, *DisruptiveCyberCrimeCase*, *CyberHarassmentCase*, *CyberTrespassCase*. The first two concepts are the application of traditional *Theft* and *Fraud* crimes to the digital domain and contain some important case concepts such as the *IdentityTheftCase*, *FinancialFraudCase*, *PhishingCase* concepts amongst others. Financial fraud is defined as those activities that require the victim to part with money in good faith for non existent good or services. Phishing occurs when a victim unwittingly parts with sensitive information that can be used later against the victim. The theft of the information is used to 'steal' a person's identity and withdraw their money or bill them for material that the attacker receives. As a result, *PhishingCase* is a specialisation of *IdentityTheftCase* and is marked as such in the ontology.

The latter three top *CrimeCase* concepts cover the *DisruptiveCrimeCase*, the *CyberHarassmentCase* and the *CyberTrespassCase* concepts. The *DisruptiveCrimeCase* defines investigations of crimes involving behaviours that disrupt regular business and includes the potentially non-legal *MisuseOfSystemsCase* concept. *CyberHarassmentCase* covers investigations of harassing or abusive behaviours, such as *CyberBullying* or *SexualHarassment*. The *CyberTrespassCase* concept defines cases of *UnauthorisedEntry* and *Hacking*. *CrackingCase* is defined as both a *HackingCase* and *DispruptiveCrimeCase* concept.

### 2.2. Information ontology

A typical computer system holds a wide variety of content. In an investigation, a small subset of relevant information that proves or disproves a criminal hypothesis is searched for. Typically the same type of information is retrieved depending on the case type. As such, an *Information* ontology, classifying different types of data, provides another dimension for describing digital forensic cases.

At the top level of the *Information* hierarchy, the sub-ontology defines the *DataObject*, the *ServiceObject* and the *SoftwareObject* concepts as the main types of information that can be found on the system. *DataObject* defines all tangible units of data in the system. The *DataFragmentObject*, encompassing such concepts as the *RegistryKeyObject* and the *PasswordObject*, is a *DataObject* and is the smallest logical unit of evidence viewed independently of any files that it may belong to. The *FileObject* is viewed as a collection of *DataFragmentObjects* rather than a single entity. Object properties *hasFileName* and *hasFileExtension* amongst others are used to identify *FileObject* individuals. Further categorisation of *FileObject* to the *MediaFileObject*, *TextualFileObject* etc., is accomplished by specialising the restrictions to specific extensions and particular metadata that defines these file types.

The *SoftwareObject* concept identifies software and applications that are found on the system. Full description of software in terms of artifacts, actions and language as in (Lando et al., 2007) is beyond the scope of the system. At the moment, we treat *SoftwareObject* as a static entity stored in the file system which can be executed to accomplish some function. As such, the ontology specialises the concepts based on the function of the software and models the software's on-disc structure by relating with relevant *DataObjects*, files and folders, belonging to that *SoftwareObject*.

The two highest specialisations of *SoftwareObject* are *PersonalApplicationSoftware Object* and *SystemSoftwareObject*. The former encapsulates all software that the user installs while the latter conceptualises the *OperatingSystemObject* concept. *UtilitySoftwareObject* is *PersonalApplicationSoftwareObject* and defines all the small tools that manage, tune and organise the data for the benefit of the user, such as anti-virus software. These small tools usually carry out a small number of tasks to indirectly benefit the user. In contrast, *ApplicationSoftwareObject* is a *PersonalApplicationSoftwareObject* installed by the user to directly create, edit or view data or execute major tasks. The majority of personal software is covered by this concept and includes *WordProcessingSoftwareObject*, *IMSoftwareObject* amongst others.

The *ServiceObject* concept is not a *SoftwareObject* nor a *DataObject* but a service provided by remote providers, such as a web site or a remote storage provider, to the user. Examples of these services are the *InternetForumSite*, the *SocialNetworkingService* amongst others. Each may leave specific evidence on the host system but is not installed into the system itself.

The *Information* ontology also contains a container concept, *EvidenceObject*, which relates to the forensic field specifically. It contains collective concepts relevant to forensics such as the *UserActivityEvidence* concept, the *SystemConfigurationEvidence* concept, the *UserProfile* concept amongst others. The concepts combine *ServiceObject*, *DataObject* and *SoftwareObject* and help identify evidence relating to a particular aspect of the investigation. For example, the *CommunicationEvidence* concept relates to the evidence of communications between the owner of the digital device and any other third party. The concept references *DataObjects*, (*EmailFileObject*), *SoftwareObjects* (*FileSharingSoftware*), and *ServiceObjects* (*Forums*) through appropriate object relations. Other evidence concepts, such as the *UserActivityEvidence* and the *GamingActivityEvidence* are defined in a similar manner. As well as describing evidence, these concepts also allow DIALOG to annotate evidence in a single case and behave as a case manager.

### 2.3. Information location ontology

Potential evidence may reside in a variety of locations. Any of the data described in the *Information* ontology in Section 2.2 can be stored in any number of different locations in the file system. The location of many important files however, such as system and application log files tend to be easily predicted. Other less structured data, such as user files, can benefit from the generality that an ontology brings and guide the investigator to the most probable location. The *InformationLocation* ontology captures this element of the investigation.

The top concepts of the *InformationLocation* are the *DigitalLocation* and the *ConventionalLocation* concept. The latter defines those locations that have relevant evidence for the investigation but are not of the digital type. These include *ReferenceMaterial* such as the *ComputerManual* and the *FilePrintouts* concepts. These concepts are analogous to the *NonCyberCrimeCase* concept in Section 2.1. Both relate more to the traditional non cyber crime element of the investigation, but are relevant and are included in the respective ontologies (U.S. Department of Justice, 2008).

The *DigitalLocation* concept defines all locations that store the information in a digital format. It differentiates between the *PhysicalLocation*, those locations that have a physical dimensions, and the *LogicalLocation*, those locations of data irrespective of the physical medium it is stored on. The former concept encompasses physical objects that store information, i.e. the *DigitalDevice*, and those units of physical space that make it possible for the information to be stored, the *LowLevelLocation* concept. The *LowLevelLocation* concept is a physical location of data that is hidden from the user but that is relevant to the forensic examiner, such as *SlackSpace*, *SwapSpace* and *FreeSpace*, collectively termed *AmbientDataLocations*.

The *DigitalDevice*, conversely, is a macro location that can store relevant digital data and is defined as an appliance used in conjunction with computers or as a computer replacement. The *SmallScaleDigitalDevice* and *LargeScaleDigitalDevice* concepts encompass the two main different types of these devices. The former can be defined as any portable device designed to carry out a limited number of digital tasks and include the *ThumbDrive*, the *Printer*, the *MobilePhone* etc. The definition is broader than those found in (Harrill and Mislan, 2007) but differentiates from the second group of devices. The *LargeScaleDigitalDevice* is a device of one or more interconnected computers designed to do or facilitate a multitude of digital tasks. These include the *Grid*, the *Server* and the *PersonalComputer* itself.

Every *DigitalDevice* references its data in a logical way to hide the physical manner that the data is stored. The logical addressing of the data is conceptualised in the *Logical Location* concept. Two types of logical address are specified, the *RemoteResource Location* and the *LocalResourceLocation*. The former concept defines those locations outside of the local *DigitalDevice* such as *IPAddress* and *WebpageAddress*. The latter, *LocalResourceLocation*, concept is the opposite. It conceptualises the location of local resources and defines the *OnDiscLocation* and the *FileSystemLocation* concept, such as the *FilePath*, *FATEntry* and *MFTEntry* etc. To facilitate addressing of specific elements within files themselves, the *IntraFileLocation* concept is also defined. The paths of specific registry keys, the location of embedded data structures, metadata amongst others are defined as being *IntraFileLocations*.

## 2.4. Forensic Resource ontology

The forensic program is the basic apparatus of the cyber crime investigation. It is used to extract, analyse, preserve and present all form of digital evidence. They provide a resource to the investigators to achieve their aims and therefore are an important dimension in describing the investigation itself. The *ForensicResource* ontology defines these resources and relates them to the relevant data locations and data that they operate on. It identifies two types of resource, the *ForensicSoftwareObject* and the *ForensicServiceObject*.

The *ForensicServiceObject* concept is a *ForensicResource* that provides assistance to investigators through the dissemination of valuable information. Typically coming in the form of a *ReferenceService*, these forensic resources include the *HashDatabaseService*, the *ReportingServiceObject* etc. DIALOG itself can be considered an instance of the *ForensicServiceObject* concept. Semantically, as well as a *ForensicResource*, the *ForensicServiceObject* is also a *ServiceObject* previously defined in the *Information* sub-ontology and is related to that ontology in the appropriate manner.

The *ForensicSoftwareObject* is similarly related to the *SoftwareObject* of the *Information* sub-ontology. However it is also a *ForensicResource* which conceptualises those software tools that can be used to carry out an investigation. The concepts follow closely the definition of main investigation stages identified in many forensic guides. Namely, the *PreparationSoftwareObject*, the *DetectionSoftwareObject*, the *Acquisition SoftwareObject*, the *EvidencePreservationSoftwareObject*, the *AnalysisSoftwareObject* and the *ReportingSoftwareObject* concept.

The *PreparationSoftwareObject* concepts defines software that are used prior to any crime ever happening. They are used to assess risk, educate personnel and train investigators for any crime that may warrant investigation in the future and include the *SurveySoftwareObject* and the *CrimeMappingSoftwareObject*. The *Detection SoftwareObject* concept, on the other hand, defines those tools that can be used to alert relevant parties of a crime occurring at that instant. They are used as a preventative measure or against a person who is suspected of committing a criminal activity. Amongst others, the concepts cover the *NetworkSnifferObject* and the *KeyLoggerObject*.

The aforementioned tools are designed to be applied proactively to stop crime from happening, the remaining categories cover those investigative tools designed to be used in the traditional reactive sense when a crime has already been suspected to have occurred. They cover the 'Acquisition' phase, *ImagingSoftwareObject* for example, the evidence 'Preservation' phase, the *HashingSoftwareObject* concept for example, the analysis phase and the reporting phase of the investigation.

The 'Analysis' phase defines the majority of tool types. Four sub-types of analysis software have been identified and are defined by the *BrowserSoftwareObject*, the *ConversionSoftwareObject*, the *FilteringSoftwareObject* and the *DataCorrelation SoftwareObject* concepts. The *BrowserSoftwareObject* defines those softwares that merely present data to be inspected, such as the *HexViewer*. The *Conversion SoftwareObject* concepts defines those softwares that convert data from one format to another. The conversion is typically from a less understandable state of data to another more understandable one. The process is reversible and verifiable. The *DecryptionSoftwareObject* concept as well as the traditional *FileFormat ConversionSoftwareObject* concept belong to this category.

The *FilteringSoftwareObject* concepts defines those softwares that take a large amount of data as input and return a smaller subset of data that has passed a certain condition specified by the user. They encompass the *KeywordSearchSoftwareObject* as well as the more complicated *PatternRecognitionSoftwareObject* concept. Typically, every investigation will involve some form of searching and many software will be applicable to this category. The final *AnalysisSoftwareObject* concept is the *DataCorrelation SoftwareObject* concept. This concept defines softwares that take a small number of disparate data and relate them to each other to highlight relevant evidence. *TimeStamp CorrelationSoftwareObject*, *FileComparerSoftwareObject* and other crime scene reconstruction software are examples of these types of tools.

## 2.5. Other ontologies

Other smaller ontologies are also utilised to define other relevant concepts of cases. In particular a small *Actor* Ontology is used to define the various parties involved in an investigation. This simple ontology only defines the *ComputerisedActor*, the *HumanActor* and the *HumanOrganisation*. The sub-ontology will be further enhanced by the inclusion of established Actor ontologies such as the *Friend Of A Friend* (FOAF) ontology (Brickley and Miller, 2007).

## 3. Modeling the registry

The concepts of DIALOG in Section 2 constitute a general description of the main parts of the investigation. The ontology will be refined down towards specialised subjects to define them in more detail. As an illustration, we will model the knowledge associated in the Windows Registry. The registry, as mentioned before, is analogous to a file system and contains a huge variety of information and is analysed in the majority of cases. The structure of the registry will be modeled with respect to both the structure and the type of evidence that the specific registry keys hold. Modeling of the entire file system can be conducted in a similar way.

### 3.1. Modeling the registry structure

The registry is a hierarchical database constructed of two main elements, the key and the value. Each key can contain one or more subkeys and is analogous to the folder in the file system. The values hold the actual data and are analogous to files. Both the value and the key are named but only the key is time stamped and contains a Last Modified Time field.

The registry combines keys stored in a number of different hive files to a single central database. Each key is referenced with a unique path using this single database perspective. The path makes no distinction where in the file system the key is stored which in reality could be in one of five main hive files. The same key with the identical name and function may exist in different places of the registry and have only a very subtle difference in their meaning. Tables 1 and 2 shows an example of a key used to store the path of documents that were accessed most recently.

Any system that attempts to define the registry key accurately must take into account these structural properties first. Logically they can be represented as axioms or rules that constrict the keys definitions. The rules are summarised below.

1) Key has Key min 0.
2) Key has Value min 0.
3) Key has LastAccessedDate exactly 1.
4) Key has RegistryPath min 0.
5) Key isIn RegistryHive min 0.

Any structure that fulfils the axioms above can be inferred to be a registry key. From an information point of view, the registry key is a small fragment of data that holds a very specific type of content. It is an instance of a *DataObject* and falls under the *DataFragment* concept in the *Information* sub-ontology. It is also a *DataContainerObject* that may hold one or more different registry values. Both are encapsulated in the *RegistryKeyObject* and *RegistryValueObject* concepts and are related to each other with *Object* relations. The DIALOG ontology also provides File and Data Location concepts to define the *RegistryHiveObject* and the *RegistryPath* components of a key. They too are related to the *RegistryKeyObject* concept with similar *Object* relations. The *Name* and *LastAccessedDate* attributes, conversely, are represented by Datatype properties as they do not have a conceptualisation in the ontology.

Fig. 2 illustrates the modeling of the structure of a key with DIALOG. All instances contain these properties upon creation. *Cardinality* and '*Necessary and Sufficient*' conditions exist to enforce the instantiation of certain essential attributes. All keys, for example, must contain a name for the key to be created.

### 3.2. Modeling the registry semantics

Each key of the registry serves a purpose in the Operating System. Since the number of keys is so vast the functions vary greatly and have different implications in the forensic investigation. Semantic modeling deals with modeling the interpretations of what the keys are designed to do rather than how they are constructed.

To model the functions of the key, the *Information* sub-ontology contains concepts specific to digital forensics evidence. The *EvidenceObject* component mentioned in Section 2.2 describes the functions of the information irrespective of what type of information that evidence is. The concepts, including *CommunicationEvidence*, *MultiMediaEvidence*, *SystemConfigurationEvidence* and others, provide a vocabulary to tag individual registry keys and other information fragments with evidence concepts related to their function.

There are two major concepts of evidence defined; the *PassiveEvidenceObject* and the *TemporalEvidenceObject* concept. The former encapsulates all evidence objects that provide evidence of an event occurring at a single point in time. The latter, conversely, provides evidence of activity ranging across a time range. The activity typically has to be inferred from one or more passive evidence objects but can range from a very short time period, such as a single user session to a longer period such as the lifetime of the computer. The user activity is vitally important to investigators and typically requires a large amount of reasoning. Further sub concepts are defined for both of these types of evidence.

The definition of a registry key can be enhanced by defining it to be both a *RegistryKeyObject* and any number of *EvidenceObject* concepts. At present, this tagging must be carried out manually when an instance of a registry key is created. However, an automated tagging system based on more accurate definitions and axioms can be developed to carry this out automatically.

As an example, we will illustrate the enrichment of the definition of the 'RecentDocs' MRU key used in Section 3.1. This MRU (Most Recently Used) key holds an ordered list of the last documents accessed by the user. The file system path of any document opened or edited by the user is entered in this key. As such it provides an important clue to user activity and is usually analysed in a forensic investigation. Here, we extend its definition to reflect this function.

The primary role of this key is to display a small list of names of the most recently opened documents in the My Recent Documents area of the Start menu. It is a point of convenience designed to improve the experience of the user. From a forensic point of view, the key reveals a number of different types of evidence. Firstly, the key entries are file name entries. They testify that the files with those names *exist* or have *existed* in the file system at some stage in the recent past. The key, also contains an ordering to that list, specifying the order at which these files were accessed by the user. As such they reveal a limited *user history* with respect to those

**Table 1 – Summary of the structure of the 'RecentDocs' key.**

Details of the 'RecentDocs' registry key

| Property | Type | Number | Example |
|---|---|---|---|
| Name | String | 1 | RecentDocs |
| LastModifiedDate | Date and Time | 1 | 02/03/2009 14:16:38 UTC |
| Subkey | Registry Key | >0 | avi, .dat, .doc, .exe, .flv, .rar, .txt, .zip |
| Values | Registry Value | >0 | MRUListEx |
| Logical Path | Registry Path | >1 | HKLM\Software\Microsoft\Windows\CurrentVersion\Explorer\RecentDocs |
| Storage Location | Hive File | >1 | SOFTWARE |

| Table 2 – Summary of the function of the 'RecentDocs' key. | | |
|---|---|---|
| Functional detail of registry keys | | |
| Key | Function | Concept |
| RecentDocs | - Functions as a registry key<br>- State that these entries (files) exist or have existed<br>- Information on the order that these entries were accessed in | *RegistryKeyObject*<br>*DocumentEvidence*<br>*DocumentActivity* |

files. The former role is encompassed by the *DocumentEvidenceObject* concept while the latter is encompassed by the *DocumentAccessedObject*. This key therefore is an example of both a *TemporalEvidenceObject* and a *PassiveEvidenceObject*. Table 1 summarises the roles of the key and the concepts in the *EvidenceObject* ontology that define those roles.

Other forensic keys can be annotated with similar evidence concepts relevant to their function. Not all keys are instances of *TemporalEvidenceObject*. Most keys are passive objects showing evidence for only one point in time. Keys, much like files, have a single timestamp updated every time the key is modified. However, they can still be annotated with the concepts from the *PassiveEvidenceObject* part of the ontology. *HKLM\Software\Adobe\Acrobat Reader\7.0* for example, simply testifies that the software Acrobat Reader exists on the file system and as well as being an instance of the *RegistryKeyObject* is also a *SoftwareEvidenceObject*.

The combination of the *TemporalEvidenceObject* and *PassiveEvidenceObject* instances form a knowledge base of registry keys which can be accessed by applications that require interpretation of the roles of the keys.

## 4. Applications of DIALOG: RPCompare

There are a number of programs employed in the forensic field to extract evidence useful to the investigation. By using DIALOG, software can adopt a more automated approach and present the results in an application independent environment. The results can therefore become a part of a larger investigative process using multiple software products to accomplish different tasks. It can serve as a single unifying structure in cataloging all sorts of evidence from a variety of different tools where outputs of one software are inputs of another.

In this section we use DIALOG to add semantics to the results from the registry analysis program RPCompare (Kahvedžić and Kechadi, 2008). RPCompare takes in as input the series of Restore Points present in a typical Windows system and compares the registry hives stored within them. Since the Restore Points are snapshots in time of the state of the system, the differences between them can highlight the activity that has occurred from one point to another. In order to do this, the DIALOG ontology is to conceptualise each of the software's concepts; its structure, inputs and outputs and will provide a set of inference rules that will mimic the reasoning process of human users. Although human interaction will not be totally removed, the ontology can be used to make the process simpler and the results more understandable.

### 4.1. RPCompare structure

RPCompare compares each key from the first hive with a corresponding key in the second hive. The hives are chronologically ordered with a period of time elapsing between the creation of one restore point and another. If a key exists in the first hive and does not exist in the second, then that key has been *Removed* during this period. Similarly, if a key is found in

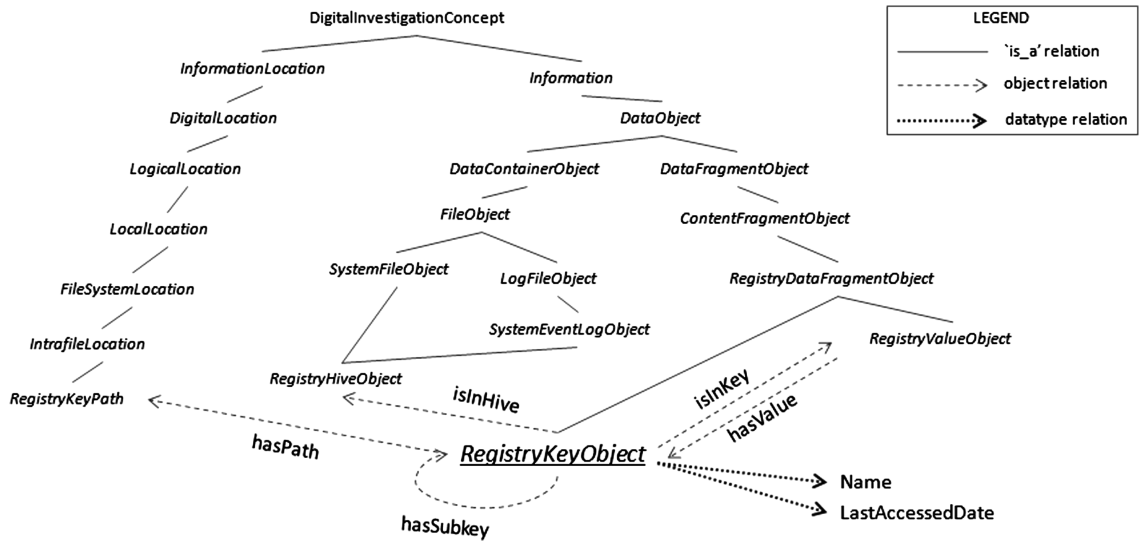

Fig. 2 – Conceptualisation of a Registry Key.

```
1ST Hive vs 2nd Hive                                                              2nd Hive vs 3rd Hive

ADDED: \Software\Adobe\Acrobat Reader                                              ADDED: \Software\Microsoft\Windows\ShellNoRoam\Bags\1875
ADDED: \Software\Adobe\Acrobat Reader\7.0                                          ADDED: \Software\Microsoft\Windows\ShellNoRoam\Bags\1875\Shell
ADDED: \Software\Adobe\Acrobat Reader\7.0\AdobeViewer                              MODIFIED: \Software\Microsoft\Windows\ShellNoRoam\BagMRU
ADDED: \Software\Adobe\Acrobat Reader\7.0\Installer                                MODIFIED: \Software\Microsoft\Windows\ShellNoRoam\BagMRU\0
ADDED: \Software\Adobe\Acrobat Reader\7.0\Installer\Migrated                       MODIFIED: \Software\Microsoft\Windows\ShellNoRoam\BagMRU\0\2
MODIFIED: \Software\Microsoft\Office\12.0\Common                                   MODIFIED: \Software\Adobe\Acrobat Reader\7.0\AdobeViewer
MODIFIED: \Software\Microsoft\Office\12.0\Common\HelpViewer                        MODIFIED: \Software\Adobe\Acrobat Reader\7.0\AdsInReader
MODIFIED: \Software\Microsoft\Windows\CurrentVersion\Explorer\ComDlg32\LastVisitedMRU   MODIFIED: \Software\Adobe\Acrobat Reader\7.0\AVGeneral
MODIFIED: \Software\Microsoft\Windows\CurrentVersion\Explorer\ComDlg32\OpenSaveMRU\*    MODIFIED: \Software\Adobe\Acrobat Reader\7.0\AVGeneral\cRecentFiles\c1
MODIFIED: \Software\Microsoft\Windows\CurrentVersion\Explorer\ComDlg32\OpenSaveMRU\zip  MODIFIED: \Software\Adobe\Acrobat Reader\7.0\AVGeneral\cRecentFiles\c2
MODIFIED: \Software\Microsoft\Windows\ShellNoRoam\BagMRU\2\3\0\27                  MODIFIED: \Software\Adobe\Acrobat Reader\7.0\Collab\cServerSettings
MODIFIED: \Software\Microsoft\Windows\ShellNoRoam\BagMRU\2\6                       MODIFIED: \Software\Adobe\Acrobat Reader\7.0\RememberedViews
MODIFIED: \Software\Microsoft\Windows\ShellNoRoam\Bags                             REM: \Software\Grisoft\Avg7
MODIFIED: \Software\Microsoft\Windows\ShellNoRoam\Bags\1018\Shell                  REM: \Software\Grisoft\Avg7\Config
MODIFIED: \Software\Microsoft\Windows\ShellNoRoam\Bags\1131\Shell                  REM: \Software\Grisoft\Avg7\Config\AvgAPI
```

Fig. 3 – Results in RPCompare.

the second hive and not in the first then that key has been *Added* in the interim. If the same key exists in both but has different content, then that key has been *Modified*. RPCompare can compare a single key, a set of keys, or the complete set of registry hives.

Therefore, the main function is to correlate similar data (the registry keys) located at different points in the files system (the Restore Point folders). As such RPCompare is an instance of the a *DataCorrelationSoftwareObject* concept in the *ForensicResource* sub-ontology. Specifically a *ComparerSoftwareObject* concept. It takes as input at least two similar data containers (keys, hives or Restore Points) and returns containers storing either removed, deleted or modified registry keys. These containers are instances of the *RPCompareContainerObject* concept and are a specialisation of the *VolatileContainerObject* concept. These concepts are summarised below. Other attributes such as software Author, Owner, Execution environment etc are omitted.

```
RPCompare isAnInstanceOf ComparerSoftwareObject
RPCompare takesAsInput
    RegistryKeyObject atleast 2 or
    RegistryHiveObject atleast 2 or
    RestorePointObject atleast 2
RPCompare returnsOutput RPCGroupObject
RPCGroupObject contains RPCUnitObject
RPCUnit contains RegistryKeyObject and
RPCUnit   hasModifiedState{''Modified'',''Removed'',
''Added''}
```

Comparisons result in a large number of differences. To make the process more efficient, the authors of RPCompare described a set of improvements to guide the investigator intelligently (Kahvedžić and Kechadi, 2008). First by using RPCompare on progressively smaller number of keys and then applying a series of simple rules to classify the resulting differences. The general methodology is user intensive and does not have any transparency and formality in the classification rules. The rules are not checked for consistency, are not used to automatically infer new knowledge and are very application specific. Here we attempt to formalise the rules by mimicking the human investigator reasoning process.

Consider Fig. 3, it shows a very small sample of differences between three user registry hives. We identify two sets of manual reasonings techniques. The first occurs if the user knows the function of the key. For example, installed software usually places its registry keys using the *HKCU\Software\Manufacturer\Product* convention. Since the *HKCU\Software\Adobe\Acrobat Reader\7.0* key was added in the first set of results it can be inferred that the software Acrobat Reader Version 7 was installed. Secondly, changes in unknown keys, such as *HKCU\Software\Adobe\Acrobat Reader\7.0\AVGeneral\cRecentFiles\c1* can be attributed to the same Adobe product because it is found under the same registry branch. Even if the role of the key is not known, it can at least be inferred that it was likely to been added by Acrobat Reader. The aim of DIALOG is to mimic this simple reasoning process.

Therefore, inferring meaning out of RPCompare results is achieved in two steps:

1) **Identify Key**: Extract the function of the removed/added/modified key.
   a) Find the system component that owns this key.
   b) Find the component's function.
2) **Infer Meaning**: Interpret the difference of the component across time.

### 4.2. Identification of RPCompare results

The first step is achieved by querying the ontology knowledge base for the particular key. If it exists then the role of the key can be directly accessed. For example, the key *HKLM\Software\Microsoft\Windows\CurrentVersion\Run* is a well known key storing the paths of the software that is executed when the system starts up. As well as a *RegistryKeyObject*, this key is also an instance of *SoftwareEvidenceObject* and *SystemStartUpEvidenceObject*. It has been identified as an important forensic key and has been inserted in the ontology knowledge base manually. The latter two concepts can therefore be directly accessed by RPComparer and presented in the results to add semantics to the key.

However, similar to the way the user cannot remember all functions of all keys, it is impractical to annotate all possible keys with the evidence objects. Some grouping must be applied. To this end, we introduce two types of *RPContainerObject* concept each of which stores particular aspects of the results. The *RPCUnitObject* instances store changes with respect to a single comparison while *RPCGroupObject* instances stores one or more similar units grouped under a common key. Each group contains one common parent key the function of which is known. Reasoning on the entire group as a whole can be achieved by reasoning on this one group key.

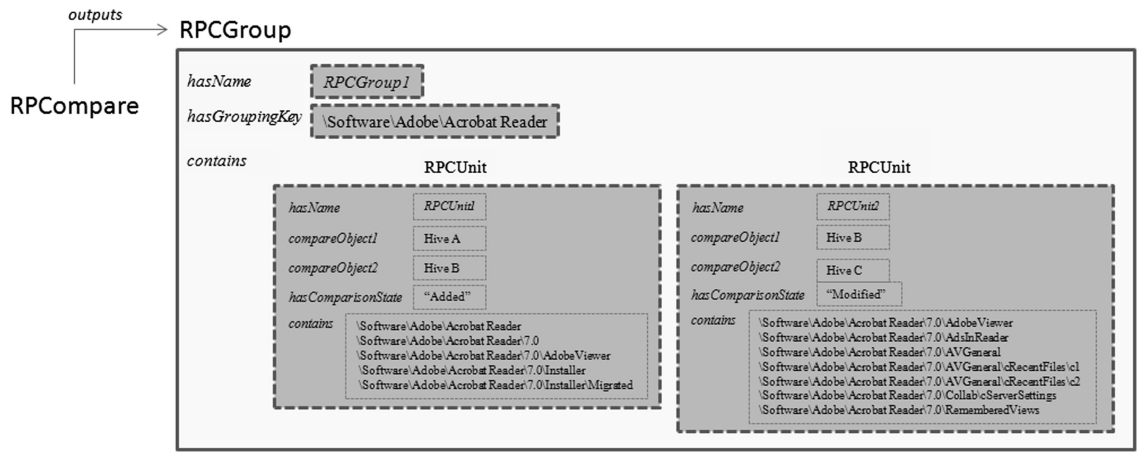

**Fig. 4 – Conceptualisation of the Results using *RPGroupObject* and *RPUnitObject* containers.**

A limited set of grouping keys have been defined. Initially RPCompare will create groups based only on the *HKCU\Software\Manufacturer\Product* convention. This will encompass any software changes since this is the convention most software follows. Other grouping keys such as *HKCU\Software\Microsoft\Windows\ShellNoRoam*, storing positional and access information for folders, *HKCU\Microsoft\Windows\CurrentVersion\Explorer\FileExts* storing file extension settings of Explorer etc. will be defined incrementally. Fig. 4 illustrates one group of keys from the results in Fig. 3. The keys in this case contain two units grouped under *HKCU\Software\Adobe\Acrobat Reader* key which has been defined in DIALOG as both a *RegistryKeyObject* and a *SoftwareEvidenceObject* concept.

Growth of the registry key knowledge base for identification can therefore be an incremental process with progressive definition of grouping keys.

### 4.3. Inference of RPCompare results

Once a key has been found to be removed/added or modified by RPCompare and its function specified in the above manner, DIALOG can be further utilised to reason about the activity of the key over a period of time. As mentioned before, the DIALOG evidence sub-ontology contains temporal evidence concepts to annotate the results returned by RPCompare. A number of specific categories of activity that RPCompare reveals most readily are summarised below. More specialised activity can be progressively defined in a similar way to the other concepts in DIALOG.

- **Software Installation/Uninstallation**: Any activity whereby a software program has been installed or removed from the file system.
- **Software Configuration**: Any activity which results in a change of configuration of a software, including the Operating System itself.
- **User File Activity**: Any evidence of activity relating to files. Reveals evidence of file access and creation.
- **User Folder Activity**: Any evidence of activity relating to folders. Usually this reveals evidence of folders only, not creation.

The inference of new knowledge and the reasoning about the individuals is accomplished using the Semantic Web Rule Language, SWRL (W3C, 2004). SWRL has been designed to provide rule functionality to ontologies and continues to receive a great amount of development. SWRL rules consist of an antecedent condition and a consequent. The consequent only being executed if the antecedent evaluates to true. It takes as arguments any concept or relation in the ontology and are used with the RPCompare concepts to infer new relations and knowledge.

For example, the following simple rule asserts that if a *RPCGroupObject* instance holds a common key, as described in the previous section, then the system component whose changes they encapsulate is the same as the software that owns the common key. In the case of the group in Fig. 3, this rule would imply that the group is evidence of Adobe-Acrobat software since the grouping key, *\Software\Adobe\Acrobat Reader*, has been asserted to belong to that software.

```
<antecedent>
   di:RPUnit(?di:x) and
   di:hasCommonKey(?di:x, ?di:y) and
   di:belongsToSoftware(?di:y, ?di:soft)
<consequent>
   di:isEvidenceOfSoftware(?di:x, ?di:soft)
```

More complicated rules can be built incrementally. The following rule states that if a *RPCUnitObject* instance has a comparison state "Added" and contains a registry key that is a child of that group's common key and that common key has a path of *HKCU\Software* then that *RPCUnitObject* is evidence of that software's installation. In the case of Fig. 3, RPCUnit1 would correctly be classified as a *SoftwareInstallationActivityObject* since it contains *\Software\Adobe\Acrobat Reader\7.0* a direct subkey of the common key *\Software\Adobe\Acrobat Reader*. Further knowledge, with respect to the installation can be added after this inference, such as what software was installed, when and to which user the software can be attributed to.

```
<antecedent>
  di:RPGroupObject(?di:obj) and
  di:isEvidenceOfSoftware(?di:obj,  ?di:software)
  and
  di:containsUnit(?di:obj, ?di:x) and
  di:RPUnitObject(?di:x) and
  di:hasComparisonState(?di:x, ''Added'') and
  di:contains(?di:x, ?di:k) and
  di:RegistryKeyObject(?di:k) and
  di:hasParentKey(?di:k, ?di:p) and
  di:RegistryKeyPath(HKCU-Software) and
  di:hasRegistryPath(?di:p, HKCU-Software)
<consequent>
  di:SoftwareInstallationActivityObject(?di:x)
  and
  di:hasSoftwareInstalled(?di:x, ?di:software)
```

These rules illustrate how the rules can be applied to the ontology to extract and infer new knowledge based on existing knowledge in the ontology. They demonstrate two types of knowledge that can be inferred. Further rules are being defined for the RPCompare software to fully embody the reasoning employed in the results. Once executed these rules will generalise the results to the evidence concepts provided by DIALOG. Therefore, once a new forensic cases is started, it can pin point the exact type of evidence required for its completion. Although RPCompare generally extracts only user activity, further softwares can also be empowered to use the ontology and provide a single and unified concepts base for all types of evidence. A validation methodology will also be developed to assert that these rules are accurate.

## 5. Discussion and future work

This paper presented a model to encapsulate the knowledge associated with digital investigation cases. It provides a vocabulary of concepts and associations in the form of an ontology to annotate as much as possible the semantics of those cases. Four main general sub-ontologies have been defined to model four areas of these cases. These include ontologies of the cyber crime type, the types of data locations, the type of data itself and the tools used to find that data. We believe that these four branches encompass the majority of cyber crime concepts, however the ontology can be enhanced with other concepts if they are deemed to be relevant.

DIALOG, like all other ontologies, is application independent it can be used for a variety of purposes. One of the main purposes of such a model is to define various properties and attributes of important forensic concepts to make their meaning understandable by the investigators. In particular we have demonstrated this by modeling the semantics of the knowledge associated with the Windows Registry. We have modeled the registry key both structurally and semantically to reflect their role in the database. The concepts provided by DIALOG were specialised where appropriate and were used to tag the instances of registry keys with the types of evidence they hold.

As an illustration, we enhanced the registry program RPCompare (Kahvedžić and Kechadi, 2008) to use the ontology. In a similar way to the registry key, we conceptualised the program by encoding both its structural aspects and its results. Sample SWRL rules were presented to show how those rules can mimic the reasoning that the human applies when analysing RPCompare results.

DIALOG can also be used as a collaborative tool that unifies the evidence found in a case. It can be used in a distributed environment where different investigators can annotate evidence to the relevant concepts in DIALOG. Different evidences from different tools can also be included in this way and ontology rules can be employed to highlight any potential inconsistencies that may arise. Future work will include expanding the concepts definition in the ontology and encoding more rules to detect these errors. DIALOG was used to annotate evidence from a single source (the registry) using a single tool (RPCompare), future development will allow integration of similar evidences from multiple sources, such as mobile phones and files systems. To that end, relevant existing ontologies that define *Actors* (Brickley and Miller, 2007) and temporal concepts (W3C, 2006), for example, will be reviewed and incorporated in DIALOG in the near future.